\documentclass{WileyMSP-template}
\usepackage{indentfirst}

\usepackage{amsmath}
\usepackage{siunitx}
\usepackage{upgreek}

\begin{document}
\pagestyle{fancy}

\title{Ultrasensitive Label-free Detection of Human CEACAM5 using a WGM Resonator Based on Thin-walled Capillary}
\maketitle

	\author{Xiaolu Cui}		
	\author{Jiyang Ma*}	
	\author{Hao Wu}
	\author{Shuoying Zhao}
	\author{Xubiao Peng}
	\author{Chuyu Li}
	\author{Ruiqi Ming}	
	\author{Qing Zhao*}

	% Affiliations: Please provide adacemic titles (Prof. or Dr.) for all authors where applicable, and include an institutional email address for all corresponding authors
	\begin{affiliations}
		X. L. Cui, Dr. J. Y. Ma, Dr. H Wu, Dr. X. B. Peng, S. Y. Zhao, Prof. Q. Zhao\\
		Center for Quantum Technology Research\\
		School of Physics\\
		Beijing Institute of Technology\\
		Beijing 100081, China\\
		Email Address: $mjy\_bear@163.com; qzhaoyuping@bit.edu.cn $

		C. Y. Li, R. Q. Ming\\
		School of Medical Technology\\
		Beijing Institute of Technology\\
		Beijing 100081, China\\
	\end{affiliations}


	\begin{abstract}
	Whispering gallery mode (WGM) laser sensors, utilizing the interaction between the in-plane evanescent field and the surface vicinity, provide enhanced sensitivity in label-free sensing for bioanalysis and disease screening. However, the unavoidably excited spiral modes resulting from the weak axial confinement and their sensing potential were overlooked. In this study, a microfluidic biosensor using the localized conical modes of an active resonator based on the thin-walled capillaries was developed,  demonstrating ultrasensitive refractive index and biomolecule detection capabilities. This sensor provides nonspecific detection of bovine serum albumin (BSA) and specific detection of Carcinoembryonic antigen-related cell adhesion molecule 5 (CEACAM5) with ultra-low detection limits, large sensing range, rapid response, and a cost-effective design, making it a promising candidate for industrial-scale production. The theoretical detection limit for CEACAM5 is as low as \SI{0.38}{ag \ mL^{-1}} (\SI{5}{z\textsc{m}}), and the sensitivity in the linear region reaches \SI{0.25}{nm \ mL \ ag^{-1}}. These results are approximately an order of magnitude higher in sensitivity than currently reported active WGM biosensors, demonstrating enormous detection potential for biomarkers.
	\end{abstract}

\setlength{\parindent}{1em}

	\section{Introduction}
	Whispering gallery mode (WGM) resonators have various advantages, including the high quality factors (Q), small mode volumes, and narrow linewidth, which can significantly enhance light-matter interactions.
In recent years, WGM has been applied in various fields, such as low threshold lasers \cite{zhao2022ultralow} and optical sensing.\cite{jiang2020whispering} In optical sensing, the application of WGM not only includes traditional substance sensing \cite{toropov2024thermo} and physical quantity sensing,\cite{chen2022highly, chai2024fano} but also biological sensing \cite{chen2019biological}.
    %近年来，WGM被应用于多个领域，比如低阈值激光器，光学调制和光学传感。在光学传感中，WGM的应用不但包括传统的物质传感，如单分子检测和物理量传感，如温度传感，PH传感，电场传感，还生物和化学传感。    
    WGM biosensors provide convenience for real-time, rapid, and label-free detection of biomolecules.\cite{fan2008sensitive, baaske2014single, brice2020whispering, vollmer2012review, zhao2022optical} Some WGM biosensors are able to detect biomolecules at the \SI{}{a \textsc{m}} level.\cite{guo2022label, reynolds2016dynamic}
%第一段，WGM检测的综述

	Most WGM resonator biosensors are passive, relying on evanescent field coupling with waveguides, prisms or tapered fibers for excitation. Although these delicate passive devices can achieve higher resolution through frequency scanning compared to fluorescence or laser spectroscopy, their coupling effect is susceptible to external perturbation. In addition, the coupling devices add complexity and cost, which may hinder their widespread adoption in biosensing applications. In contrast, active WGM resonators, integrated with gain media, enable simultaneous remote laser excitation and signal readout. The active resonators are typically fabricated with gain medium either doped within the cavity material or adsorbed onto the cavity surface.\cite{li2008optofluidic, guo2020hyperboloid} However, fixed gain medium can encounter issues such as photobleaching and photothermal effects, which can reduce service time and perturb sensing performance. \cite{yang2022review} Moreover, the interaction between the gain medium and the microcavity materials or the sample solutions may also occur, further complicating the sensing process.\cite{guo2020hyperboloid, rowland2013fluorescent}
	%第二段，有源谐振器的优势
    
	Thin-walled glass capillaries are a promising approach for the development of optofluidic biosensors.\cite{yuan2023microtubule, niu2021fiber, meldrum2014capillary} Capillary optofluidic lasers integrated with microfluidic channels have the advantages of low cost, low sample consumption, high sensitivity, high signal-to-noise ratio, and narrow linewidth of optofluidic lasers. The silica wall exhibits excellent chemical and physical stability: it is greatly resistant to common organic solvents, does no swell or corrode, and possesses high hardness and resistance to heat and pressure. The gain medium and the detection solution are separated into two microfluidic channels to avoid their interaction. Therefore, toxic, bio-incompatible materials can also be appled as gain medium and its solvent.  The dye solution is continuously extracted through the capillary, eliminating the potential bleaching and heating effects from the sensing site. 
    %毛细管在光流控上的优势
    
	Compared with microspheres, microbubbles and microbottles, capillary structure exhibits fewer variations along the cylinder axis, which results in diminished mode competition and a simpler, more easily identifiable spectrum. The cylindrical structure provides numerous sites along the axis that can be utilized for pumping and sensing, enabling the detection of multiple biomarkers within a single sample by applying various surface modifications. The simple low-cost fabrication and preparation process makes this capillary structure highly suitable for mass production in sensing applications.\cite{yang2020mass}
	%薄壁小尺寸毛细管的优势
    
	Capillary tubes cannot maintain uniform wall thickness and radius along the longitudinal direction, especially after being melted and stretched. When the wall thickness approaches the radial dimension of the first-order WGM in-plane distribution, the light field exists across both bondaries of the three layers with different refractive indices. \cite{guigot2024classification} Hence slight wall thickness changes can cause fluctuations in the effective refractive index along the axial direction. The potential energy distribution along the axial direction $V (z) \propto - r_{eff} (z)$ has two basic structures that can host laser emission: the microbottle structure with effective radius tapered at both sides, and the conical structure composed of a short section of the capillary with small and slow linear effective radius variation and a defect scattering point. \cite{sumetsky2011localization} The conical structure can host a strongly localized state on the narrower side of the defect scattering point because of the destructive interference of the wave turning from the narrower side and the wave scattered from the defect point, called a conical mode. \cite{sumetsky2011surface,poon1998spiral,lock1997morphology} The conical modes have unique response characteristics to perturbation, such as change in refractive index or protein molecule concentration, compared to the microbottle modes. \cite{cheeney2020whispering, lane2014whispering} The perturbation is based on the modification of the conical structure profile, results in a change in the effective slope, causing the resonance wavelength of the laser spectrum to shift blue with increasing refractive index or protein concentration. \cite{zhao2020dynamic}
	%阐述圆锥模式
    
This paper proposes an ultrasensitive inexpensive thin-walled capillary resonator to measure the bulk refractive index and to detecte biomolecules such as bovine serum albumin (BSA) and Carcinoembryonic antigen-related cell adhesion molecule 5 (CEACAM5) in aqueous medium. The design leverages combination of an active laser resonator and a conical liquid core optofluidic device.  The conical WGMs exist in the narrower conical capillary region next to the defect scattering point in the thin capillary wall. The laser responds to the concentration of BSA from \SI{10}{zg \ mL^{-1}} to \SI{10}{ng \ mL^{-1}} in nonspecific detection and to the concentration of CEACAM5 from \SI{100}{zg \ mL^{-1}} to \SI{1}{fg \ mL^{-1}} in specific detection, showing great sensitivity to a slight change in the concentration of biomolecules and promising potential in early cancer screening.

	\section{Results and Discussion}
	\subsection{Optical Characteristics}
\textbf{Figure 1a} illustrates the schematic diagram of a capillary laser sensor. The longitudinal variation of the effective radius generates two distinct localized modes: microbottle modes and conical modes. Figure 1b depicts the sensor's packaging architecture. The fabricated thin-walled capillary is hermetically sealed within a microfluidic chip, with a high refractive index gain medium infused into the inner channel, and surface-functionalization reagents/test solutions loaded in the outer fluidic compartments. Figure 1c shows the longitudinal structure of the tapered cappilary. The pulsed laser (wavelength \SI{532}{nm}) is focused on a short segment of the capillary, pumping the gain medium within the inner channel. The outer wall traps light by total internal reflection, sensing the change in the evanescent field. The pumped microlaser is shown in Figure 1d. The microbottle modes generate directionalless laser emission along the tangent direction around the capillary outer boundary, while the conical mode generates directional laser emission mainly from the defect scattering point by a small angle away from the tangent direction. \cite{poon1998spiral}

The capillary is fabricated through hydrofluoric acid etching, then melting and drawing in flame. \cite{zamora2007refractive} The diameter and thickness of the capillary are approximately \SI{40}{} and \SI{1}{\upmu m}. The melting process decreases the outer boundary roughness caused by acid etching. After fabrication, the capillary was fixed with ultraviolet (UV) adhesive in the groove of the polydimethylsiloxane (PDMS) microfluidic chip. The chip with the capillary was treated in oxygen plasma for \SI{2}{min} to bind with a clean glass slide, forming an enclosed microfluidic channel, where surface functionalization and sensing are performed. The relatively small diameter, although reducing the Q factor, can increase the free spectral ranges and distribute the laser energy in fewer modes, reducing the effect of mode competition on sensing stability. By changing the dye molecule concentration and the refractive index of the dye solution, the laser emission can be effectively adjusted to achieve the maximum detection range and sensitivity.

\begin{figure}
	\includegraphics[width=\linewidth]{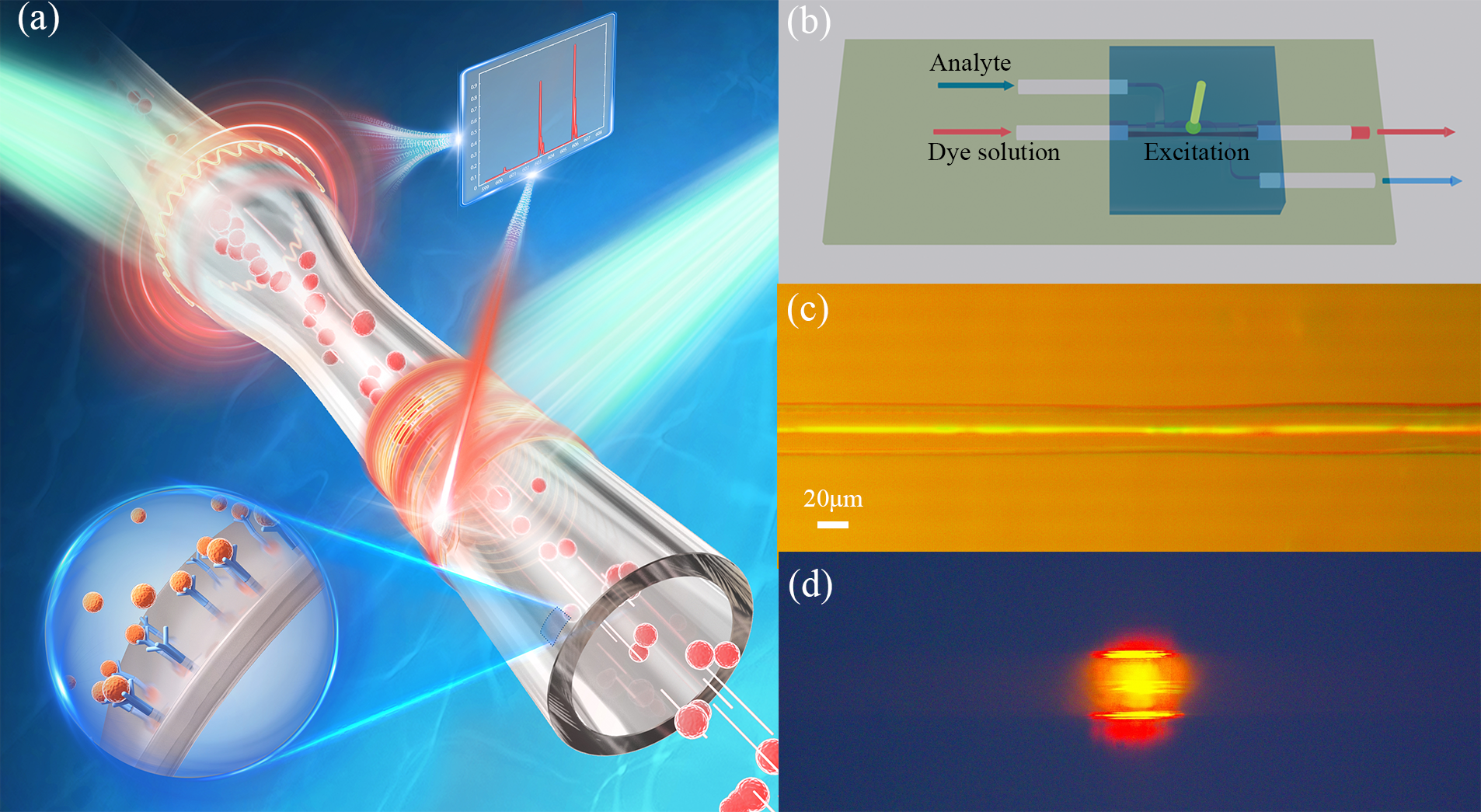}
	\caption{(a) schematic diagram of the laser biosensor. (b) Schematic diagram of the capillary resonator integrated with a microfluidic channel. (c) Microscope photo of the tapered thin-walled capillary. (d) Microscope photo of the pumped section of laser.}
    \end{figure}

	\textbf{Figure 2a} displays the laser characteristics obtained by stimulating the capillary immersed in water with different pump energy densities. The laser threshold of the resonator immersed in aqueous solution is approximately \SI{126}{\upmu J \ mm^{-2}}. The central wavelengths of the laser peaks were obtained by Lorentz fitting. Figure 2b shows the drift of the laser peak in frequent pumping during \SI{500}{s} in phosphate buffered saline (PBS). The standard deviation of the central wavelength is \SI{1.01}{pm}. The random slight shift indicates that the capillary laser is not affected by the photobleaching and photothermal effect. The stability of the sensor is sufficient for subsequent experiments.

	\subsection{Measurement of Bulk Refractive Index Sensitivity}

    \begin{figure}
	\includegraphics[width=\linewidth]{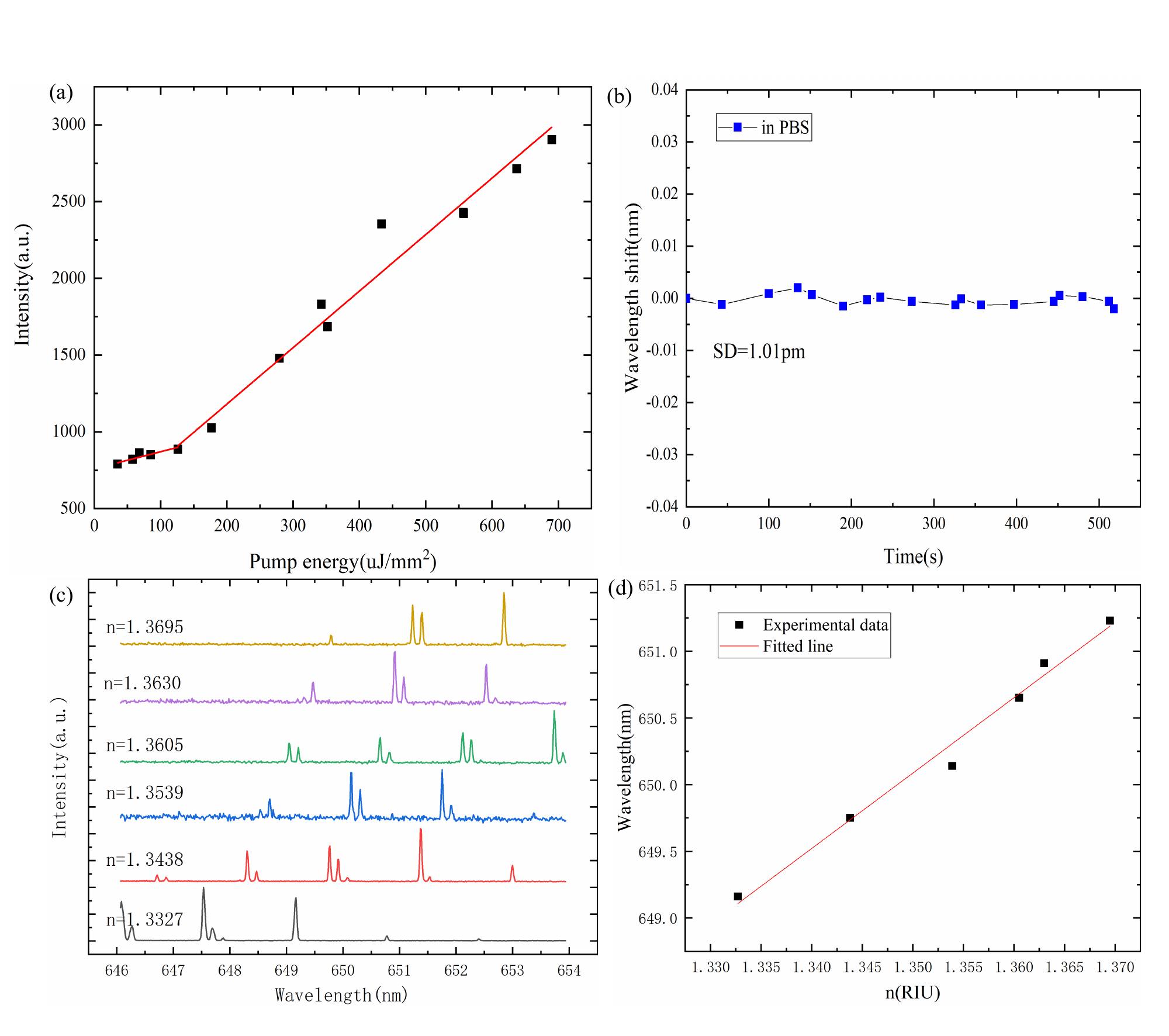}
	\caption{(a) Relationship between the laser energy and pump intensity density of a capillary resonator immersed in an aqueous solution. (b) Peak position drift obtained by multiple pumping measurements within \SI{500}{s} in PBS. The standard deviation is \SI{1.01}{pm}. (c) Spectra as a function of different refractive indices. (d) Relationship between the laser peak position and the refractive index of the solution in the microfluidic channel.} 
\end{figure}

	Aqueous solutions with different refractive indices were extracted sequentially and independently into and from the microfluidic channel using a syringe for bulk refractive index measurement. The refractive index of the ethanol solution is approximately proportional to the volume ratio of ethanol/water, can be calculated using the mixing rules and measured with a refractometer. \cite{jimenez2009concentration} The increase in refractive index caused the laser peaks of the microbottle modes to shift toward a longer wavelength. The red shift has a sensitivity of \SI{56.4}{nm \ RIU^{-1}}, as shown in Figure 2 c-d. 

For the microbottle modes, the relationship between sensitivity of refractive index and the average light field energy portion in test solution $f_3$ is \cite{zhu2007analysis}
\begin{equation}
    S=\frac{d\lambda}{dn_3}\approx \frac{\lambda}{n_{eff}} f_3
\end{equation}
where $n_{eff}$ is the average effective refractive index, $n_3$ is the refractive index of the test solution, $\lambda$ is the wavelength of the WGM mode. The measured microbottle mode refractive index sensitivity indicites that for similar diameter and thickness, the average energy portion of evanescent field is about 13\%. The perturbation caused by protein adsorption on the outer surface to the evanescent field can be estimated as an equivalent change in refractive index of the test solution. \cite{zhu2007analysis, teraoka2003perturbation, teraoka2006theory}
\begin{equation}
    \delta m_3 \approx  \frac{\alpha_{ex} n_{eff} S}{2 \epsilon_0 L \lambda n_3^2} \cdot \delta \sigma_p
\end{equation}
where $\alpha_{ex}$ is the biomolecule's excess polarizability, $\epsilon_0$ is the vacuum permittivity, $\sigma_p$ is the biomolecule surface density, $L$ is the longitudinal length of the conical mode and $S$ is the refractive index sensitivity measured in experiment.

	\subsection{Nonspecific Detection of BSA}
	The biofunctionalization and sensing process are shown in \textbf{Figure 3a-c}.\cite{kim2013protein} At first, the capillary outer boundary was immersed in $5\%$ (3-Aminopropyl)triethoxysilane (APTES) ethanol solution for \SI{2}{hours} to achieve surface silanization and then rinsed with deionized water. In this state, the capillary resonator can be used for nonspecific detection of bovine serum albumin (BSA).

	\begin{figure}
		\includegraphics[width=\linewidth]{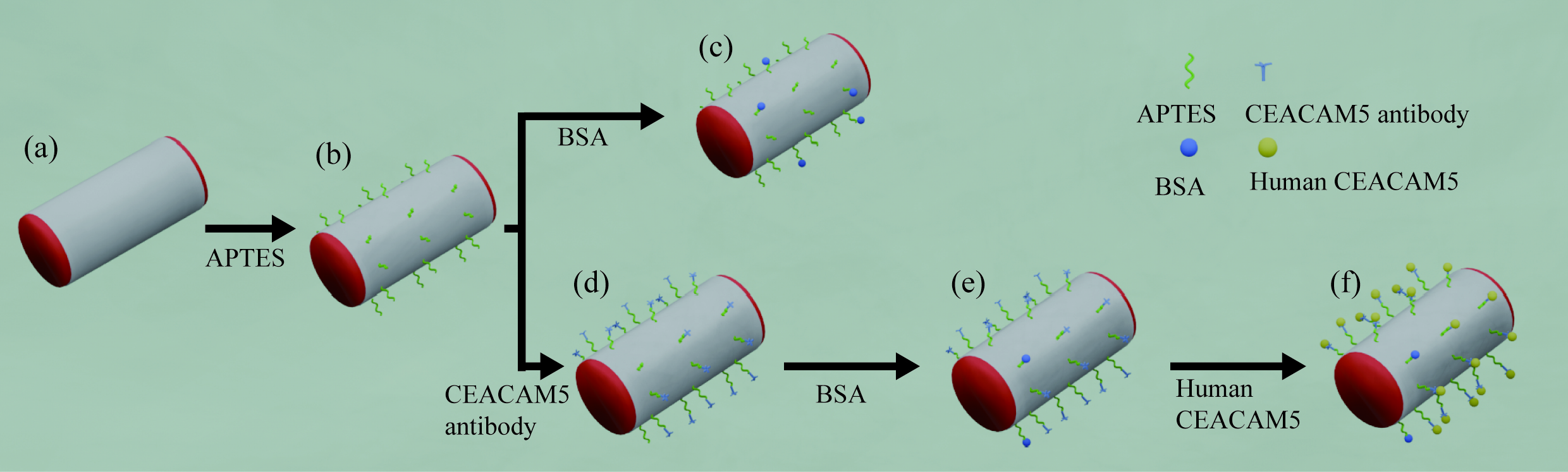}
		\caption{Nonspecific detection of BSA and specific detection of CEACAM5.} 
	\end{figure}

The binding of BSA molecules with the outer surface of the capillary demonstrated two distinct phenomena for different pumped spots, with the red shift result for the microbottle modes and the blue shift result for the conical modes, as shown in \textbf{Figure 4}. The red shift result for microbottle modes is shown in \textbf{Figure S1}.
The molecular surface density at different spots on the surface is similar, thus the average number of the detected proteins is close if the absorbed proteins are numerous. However, at similar lifetime, the photons in the conical modes traverse larger surface area, therefore having a greater chance of detecting a randomly distributed protein molecule if the absorbed proteins are rare, compared to the photons in the microbottle modes, obtaining lower detection limit.

    The accumulated wavelength shift related to the BSA concentration was fitted with Hill model ($R^2=0.995$). The sensitivity in the linear region is $S=$\SI{0.253}{nm \ mL \ ag^{-1} }. The spectral resolution of the used spectrometer is \SI{0.0135}{nm}, which was used as the standard deviation of the 3$\sigma$ method. The theoretical detection limit can be estimated as 3 $\sigma / S$=\SI{0.16}{ag \ mL^{-1}} (\SI{2.41e-21}{\textsc{m}}).
	%BSA在1ng/mL的浓度下达到相对饱和，最低检测浓度为1zg/mL(1.5x10^(-23)M)。使用Hill模型拟合了与BSA浓度和表面分子密度相关的累计波长偏移（R^2=0.995）。线性区域的灵敏度为0.253nm/(ag/mL)。使用的光谱仪的光谱分辨率为0.0135nm，用作3σ方法的标准差，则理论检测限可以被估算为3σ/S=0.16ag/mL(2.41x10^-21M)。
	
\begin{figure}
		\includegraphics[width=\linewidth]{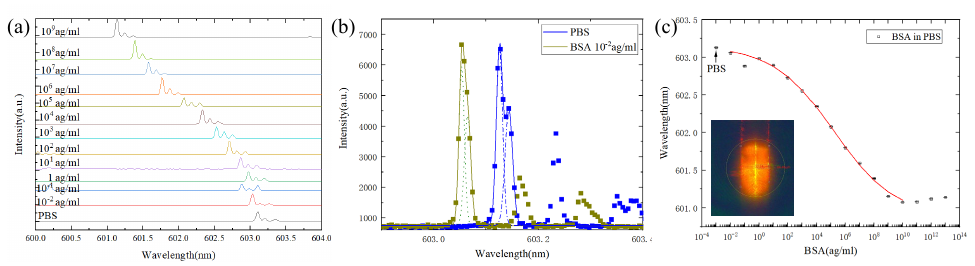}
		\caption{(a) The laser spectra as a function of different BSA concentrations. (b)The peak position shift caused by \SI{1e-2}{ag \ mL^{-1}} BSA solution. (c) The relationship between the wavelength shift, corresponding molecular surface density, and the BSA concentration. The data was fitted by Hill model ($R^2=0.995$). Insert: microscope image of the  resonator being pumped.} 
	\end{figure}

We attribute the phenomenon of blue shift in the conical mode wavelength with increasing concentration of BSA molecules to the changes in the capillary profile, that is, changes in the distribution of the effective radius $r_{eff}=n_{eff} r$ along the axis. The distribution is quantified by the effective slope $\gamma_ {eff}=| dr_  {eff} |/dz$. Along the longitudinal direction, the radial optical field at positions with small wall thickness has a larger proportion in the high refractive index gain media and evanescent fields. Therefore, having a larger effective refractive index, the effective refractive index $n_{eff} \approx f_1 n_1+f_2 n_2+f_3 n_3$ varies more with the change rate $f_3$ of the outer refractive index $n_3$. \cite{meldrum2014capillary}

The conical mode satisfies the phase self interference condition at the defect scattering point, forming a discrete resonance wavelength related to the longitudinal modulus q \cite{sumetsky2011surface, lock1997morphology}:
\begin{equation}
    \lambda_q \approx -\frac{\lambda_0}{2}(\frac{3\pi}{4}+3\pi q)^{\frac{2}{3}}(\frac{\gamma_{eff}}{\beta_0 \bar{r}})^{\frac{2}{3}}+\lambda_0
\end{equation}
where $\lambda_0$ is the in-plane wavelength, $\beta_0$ is the in-plane propagation coefficient, $\gamma_{eff}$ is the effective radius slope, $\bar{r}$ is the average radius in the conical mode range.

When BSA binds to APTES, it leads to an equivalent increase in the refractive index of the evanescent field outside the capillary, resulting in an increase in the effective slope of the mode axis within the optical field range.
\begin{equation}
    \delta \gamma_{eff}=\delta |\frac{ dr_{eff}}{dz}|=\frac{\delta(r_{eff}^e-r_{eff}^t)}{|z_e-z_t|} \approx \frac{(f_3^e-f_3^t) \bar{r}}{L}\delta n_3
\end{equation}
where  $f_3 ^ e$ and $f_3 ^ t$ are the  ratio of the evanescent field energy of the conical mode at the defect scattering point $z_e$ and the  light turning point at the cone narrow side $z_t$, $\bar{r}$ is the average cross-sectional radius within the longitudinal light field range $L=z_e-z_t$, and $\delta n_3$ is the equivalent refractive index change of the evanescent field caused by protein adsorption on the capillary surface. Therefore, the relationship between the wavelength change of cone resonance and the external refractive index change can be derived as
\begin{equation}
    \delta \lambda_q \approx -\frac{\lambda_0}{2}(\frac{3\pi}{4}+3\pi q)^{\frac{2}{3}}(\frac{\gamma_{eff0}}{\beta_0 \bar{r}}+\frac{\alpha_{ex} n_{eff} S}{2\beta_0 \epsilon_0 L^2 \lambda n_3^2} \cdot \delta \sigma_p)^{\frac{2}{3}}
\end{equation}
The formula shows that central wavelengths of the conical modes shift blue when the adsorption of protein molecules on the capillary outer wall causes the increase of the effective conical slope and the molecular surface density.

	\subsection{Specific Detection of human CEACAM5}
	Identifying specific biomarkers is a crucial approach for evaluating health or vaccination status and screening diseases in modern medical diagnosis.\cite{niu2021fiber} CEACAM5 (\SI{77}{kDa}), associated with adhesion and invasion, is overexpressed in many cancers. As a broad-spectrum tumor marker, CEACAM5 can be used for the observation of postoperative efficacy and prognosis of malignant tumors.\cite{hu2010ultrasensitive, blumenthal2005inhibition, blumenthal2007expression} In this study, a thin-walled capillary laser biosensor was used to specifically detect human CEACAM5. The surface activation and functionalization processes are shown in Figure 3a-b,d-e.
	
	Specific detection is based on the specific binding effect of mouse anti-human CEACAM5 antibodies to human CEACAM5. The CEACAM5 antibodies in PBS of  \SI{1}{\upmu g \ mL^{-1}} were introduced into the microfluidic channel and incubated for \SI{2}{hours}, where they were adsorbed on the APTES sites on the capillary outer surface, functioning as probes to capture their corresponding target protein human CEACAM5. After the antibody solution was extracted out and the channel was rinsed with PBS, BSA solution (\SI{1}{mg \ mL^{-1}}) was introduced in to block the unbound APTES sites in order to reduce the nonspecific binding of the analytes on the biosensor. \SI{30}{minutes} later, the channel was rinsed again with PBS solution to remove BSA molecules from the channel. 
	
	CEACAM5 PBS solutions of varying concentrations were introduced sequentially into the channel and incubated for \SI{20}{minutes} to reach the relatively balance state between dissociation and association at the surface sensing spots. The binding of the CEACAM5 on the conical spots results in the blue shift with great sensitivity, as shown in \textbf{Figure 5 a-b}.
	
	Figure 5 shows the relationship between the wavelength shift with the concentration of CEACAM5 in PBS. This relationship is fitted with the Hill model ($R^2$=0.987). The sensor has a relatively saturated concentration at \SI{1}{fg \ mL^{-1}}, and a lowest measurable concentration of \SI{100}{zg \ mL^{-1}}(\SI{1.30e-21}{\textsc{m}}). The sensitivity in the linear region is $S=\SI{0.106}{nm \ mL \ ag^{-1}}$. The theoretical detection limit can be estimated as $DL=3\sigma/S=\SI{0.38}{ag \ mL^{-1}}$ (\SI{4.94e-21}{\textsc{m}}). The red shift result for microbottle modes is shown in \textbf{Figure S2}.
    
    The binding of APTES and BSA occurs through amino binding. The numerous free amino groups on the surface of a BSA molecule can rapidly form bonds with the abundant APTES molecules that are situated around the outer boundary of the capillary. In contrast, the binding of mouse anti-human CEACAM5 antibodies and human CEACAM5 only occurs at relatively fewer corresponding spots on the protein structure. The antibody solution contains a large amount of BSA as stabilizer, therefore the antibody protein molecules did not occupy the entire outer boundary of the sensor. Consequently, human CEACAM5 bonds to the sensor at a relatively slower rate and reaches a lower saturation concentration. The half-maximal saturation concentration of CEACAM5 sensing is $10^5$ times lower than that of BSA, while the sensitivity is reduced only by half.  The detection range and sensitivity can be further improved  by using antibody solutions without BSA as stabilizer and of higher concentrations in the functionalization process. When the density of the antibody attached on the outer surface of the sensor is reduced, the saturation concentration would decrease rapidly, while the sensing sensitivity and detection limit would only change slightly. This is beneficial to maintain the stability of the sensing property at extremely low concentrations.
	
	In comparison to the target analyte, BSA and alpha-fetoprotein (AFP) were used as control samples to verify the specificity of the biosensor, as shown in Figure 5c. The concentrations of the tested samples were \SI{10}{ag \ mL^{-1}} and \SI{1000}{ag \ mL^{-1}}, respectively. Despite being the same concentration, the wavelength shifts caused by BSA and AFP solutions were much smaller than those caused by human CEACAM5, indicating the excellent specificity of the biosensor for biomolecular detection.
	
		\begin{figure}
	\includegraphics[width=\linewidth]{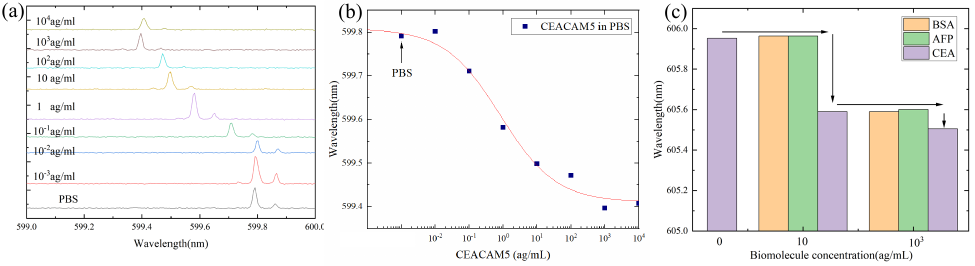}
	\caption{(a) The laser spectra as a function of different CEACAM5 concentrations. (b) The relationship between wavelength shift, corresponding molecular surface density and CEACAM5 concentration. The data were fitted by Hill model ($R^2$=0.987). (c)The wavelength shift caused by sequentially introducing PBS, and BSA, AFP, CEACAM5 of \SI{10}{ag \ mL^{-1}} and \SI{1000}{ag \ mL^{-1}}.} \
\end{figure}

\textbf{table 1} shows the different WGM biomolecule sensing methods reported in recent years. Compared with the most current optical biosensor, the thin-walled capillary laser sensor has broader sensing range and lower detection limit.
	%对比列表
	\begin{table}
		\caption{Comparison of this work and other recent demonstrated WGM biosensors}
		\begin{tabular}[htbp]{@{}lllllll@{}}
			\hline
Platform         & Type   & Analyte   & Measure range      &DL     & Reference \\
			\hline

Silicon intride microdisk&Active&Streptavidin-biotin &10[n\textsc{m}]-10000[n\textsc{m}]&6.7[n\textsc{m}]&\cite{kim2019chip}\\
		&&complex&&&&\\

Dye-doped polystyrene&Active&Neutravidin&25[n\textsc{m}]-400[n\textsc{m}]& &\cite{reynolds2016dynamic}
		  \\
microspheres&&&&&&\\

Hyperboloid-drum & Active & Human IgG & 0.007[a\textsc{m}]-0.667[a\textsc{m}]&0.06[a\textsc{m}]        & \cite{guo2020hyperboloid} \\
microdisk        &        &           &                                   &                & \\

Hollow optical  fiber &Active  &IgG        &0.12[n\textsc{m}]-1200[n\textsc{m}]          &11[n\textsc{m}]          &\cite{yang2020mass} \\
			
Liquid crystal-amplified      &Passive &Biotin     &0 - 0.1 [\SI{}{pg \ mL^{-1}}]   &0.4[f\textsc{m}]                  &\cite{wang2023highly}\\
\\
microbubble
\\

Microbubble in &Passive&cTnl&0[\SI{}{ng\ mL^{-1}}]-2[\SI{}{ng\ mL^{-1}}]&0.59[\SI{}{ng\ mL^{-1}}]&\cite{niu2022fiber}\\
hollow fiber&&&&&&\\
		
Liquid crystal&Active&
		Acetylcholinesterase& &0.1[\SI{}{pg\ mL^{-1}}]&\cite{duan2020detection}	\\
microdroplet&&&&&&\\
		
Hollow optical fiber&Active&Horseradish peroxidase&14[p\textsc{m}]-224[p\textsc{m}]	&14[p\textsc{m}]&\cite{gong2018distributed} \\

Liquid crystal-amplified&Passive&Vanillin&5-300[\SI{}{\SI{}{ng \ mL^{-1}}}]&5[\SI{}{ng \ mL^{-1}}]&\cite{hao2025orientational}
\\
microbubble&&&&&&\\

Hollow optical fiber&Active&Hemoglobin&1[n\textsc{m}] - 1[\SI{}{\micro \textsc{m}}] & 0.7[n\textsc{m}] & \cite{zhang2023optical}
\\

Suspended core fiber& Passive&DNA&1.5[n\textsc{m}] - 40[n\textsc{m}] & 1.5[n\textsc{m}]& \cite{li2023label}
\\

Hollow optical fiber&Active& CEACAM5&1.3[z\textsc{m}]-13[a\textsc{m}]&4.94[z\textsc{m}]&This work	\\

			\hline
            
		\end{tabular}
	\end{table}

\section{Conclusion}
This paper presents a highly sensitive optofluidic laser biosensor employing the conical modes in the thin-walled capillaries. The biosensor adopts a structure of internal gain medium filling and outer wall functionalization. It was found that the conical modes reduce the detection limit and exhibit unique sensing characteristics. The WGM spectra responses to BSA at concentration ranging from \SI{10}{zg \ mL^{-1}} to \SI{1}{ng \ mL^{-1}} and to human CEACAM5 at concentration ranging from \SI{100}{zg \ mL^{-1}} to \SI{1}{fg \ mL^{-1}}, and the detection limit can be as low as zeptomole-level. It has also been confirmed that this biosensor has good selectivity for the CEACAM5 detection due to the high specificity between antigen and antibody. This thin-walled capillary laser biosensor is capable of detecting various types of protein molecules and models of other receptor-ligand binding, providing a high-sensitivity, fast-response and low-cost solution for the label-free protein detection.

\section{Experimental Section}
\subsection{Materials and Reagent Preparation}
The dye is rhodamine B (RhB, Sangon Biotech (Shanghai) Co., Ltd., Shanghai, China). The dye solvent is a mixture of quinoline (Shanghai Yien Chemical Technology Co. Ltd., Shanghai, China) and dimethyl sulfoxide (DMSO, Shanghai Yien Chemical Technology Co. Ltd., Shanghai, China) with a volume ratio of 60\% and a refractive index of $1.56$. APTES (Shanghai Yien Chemical Technology Co. Ltd., Shanghai, China) is mixed with ethanol at a volume ratio of 5\% for silanization of the capillary outer surface. BSA is provided by the School of Life Sciences, Beijing Institute of Technology. Recombinant human CEACAM5, mouse anti-human CEACAM5 and other antigen-antibody pairs are purchased from Shanghai Yaji Biotechnology Co.,Ltd. (Shanghai, China) The test proteins are diluted with PBS (Beijing Solarbio Science \& Technology Co.,Ltd., Beijing, China) solution to form gradient solutions ranging from \SI{1}{zg \ mL^{-1}} to \SI{10}{ng \ mL^{-1}} in powers of 10.
	
\subsection{Fabrication Process}	

\begin{figure}
	\includegraphics[width=\linewidth]{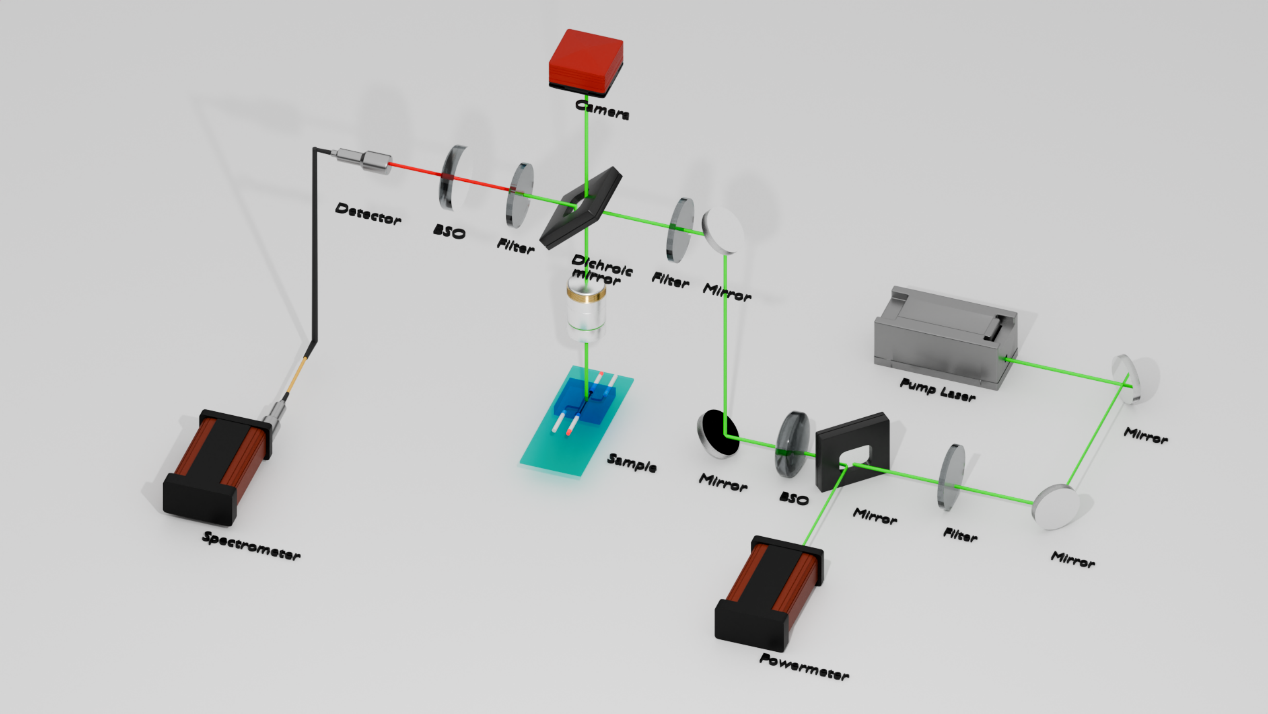}
	\caption{Experimental setup schematic of the thin-walled capillary biosensor} 
\end{figure}

The fabrication of the thin-walled capillary is similar to the reference literature. \cite{zamora2007refractive, niu2021fiber}  The fused silica capillary (TSP100140, Polymicro) is etched with hydrofluoric acid. The internal/external diameters are changed from \SI{100}{\micro\meter}/\SI{140}{\micro\meter} to \SI{100}{\micro\meter}/\SI{107}{\micro\meter}. Then, it is heated and tapered in oxyhydrogen flame to external diameter \SI{30}{\upmu m} to obtain a cylindrical thin-walled cavity with a wall thickness of approximately \SI{1}{\upmu m}. The melting and stretching reduces the roughness caused by the outer surface etching. The capillary is fixed with UV adhesive in the microfluidic channel of a PDMS leaf, which is molded from a metal mold, the manufacting process could be refered to ref. \cite{niu2024hollow}. The microfluidic channel and the glass slide are treated with a plasma surface cleaner for \SI{2}{minutes} to be bonded, and finally connected and sealed with plastic pipes and UV adhesive. Biofunctionalization is performed inside the microfluidic channel, on the outer surface of the capillary.

\subsection{Experimental setup}
The experimental setup is shown in \textbf{Figure 6}. The pump laser is a nanosecond pulse laser (532nm, 10Hz, Dalian Institute of Chemical Physics, China), and the laser spectrum is collected with a monochromator (HRS-750, Teledyne Princeton Instruments, USA). Optical beam shaping devices are used to control the diameter and position of the pump laser. Variable attenuators are used to adjust the power of the pump laser. The laser beam is coupled into a microscope system and further concentrated to a diameter of about \SI{60}{\upmu m} under a 20x objective lens. A 3D displacement stage moves the sample to the laser spot, and the output laser is collected by an optical fiber, transmitted to the monochromator, and detected by a detector (ProEM-HRS:512, Teledyne Princeton Instruments, USA). The spectral resolution of the spectrometer is about \SI{0.0135}{nm}. A syringe pump continuously extracts the dye solution through the inner channel of the capillary, and a syringe was used to extract the solution into or out of the microfluidic channel. All experiments are conducted in an indoor environment (22°C and 40\% humidity).

	% References
	\medskip

	\bibliographystyle{MSP}
	\bibliography{ref.bib}

    	\medskip
        \appendix
	\textbf{Supporting Information} \par
    \section{Theoretical derivation}

Thin-walled capillaries naturally form structures with slow changes in radius and wall thickness along the axial direction during the melting and drawing process. In stable whispering-gallery modes, the effective refractive index of light propagation is influenced by diameter, wall thickness, and refractive index of internal and external solutions, thus is a function of axial position. Within the irradiation area of the pump laser spot, the capillary tube has two shapes that can generate weak axial confinement, which could improve the quality factor and reduce the threshold: (1) microbottle; (2) the combination of monotonically increasing effective radius and defect points on fiber that generate additional scattering dissipation. The former generates WGMs with different axial orders, where the light field is concentrated on the belly of the microbottle and scattered tangentially along the circumference, while the latter generates a conical mode confined between the defect point and a turning point, where the light field is concentrated on the narrow side beside the defect point and scattered outward from the defect point. The axial distribution of the conical mode satisfies the Schrödinger-like equation:
\[
\begin{aligned}
&\frac{d^{2}A}{dz^2} + \beta^2(z)A = C   &  \beta^2(z) &= E(\lambda)-V(z) \\
&E(\lambda) = -2\beta_0^2(\lambda_{res})\frac{\lambda-\lambda_{res}-i\gamma_{res}}{\lambda_{res}}    &  V(z) &= -2\beta_0^2(\lambda_{res})\frac{\delta r_{eff}}{r_{eff}}
\end{aligned}
\] 
Where $A(z)$ is the WGM field distributed along the axial direction, the effective energy $E(\lambda)$ is directly proportional to the wavelength change, and the effective potential energy $V(z)$ is inversely proportional to the effective radius change. The constant C represents the center energy of the pump laser range. The in-plane propagation coefficient $\beta_0$ is only related to the effective radius near the defect scattering point. $\lambda_{res}$ is the resonance wavelength while $\gamma_{res}$ is the resonance width near the defect scattering point. The conical structure can be regarded as a wedge-shaped potential well. The core idea of conical mode sensing is to utilize the sensitivity of conical mode to extremely small changes in the effective radius along the capillary axis in the potential well.

\section{Redshift data of biomolecule detection with microbottle modes}
Under the same concentration changes of biomolecules, the wavelength shift amplitude generated by the microbottle modes is two orders of magnitude smaller than that of the conical mode, thus having lower sensitivity and a higher detection limit, verifying the superiority of the conical mode in low concentration sensing.

\begin{figure}
	\includegraphics[width=\linewidth]{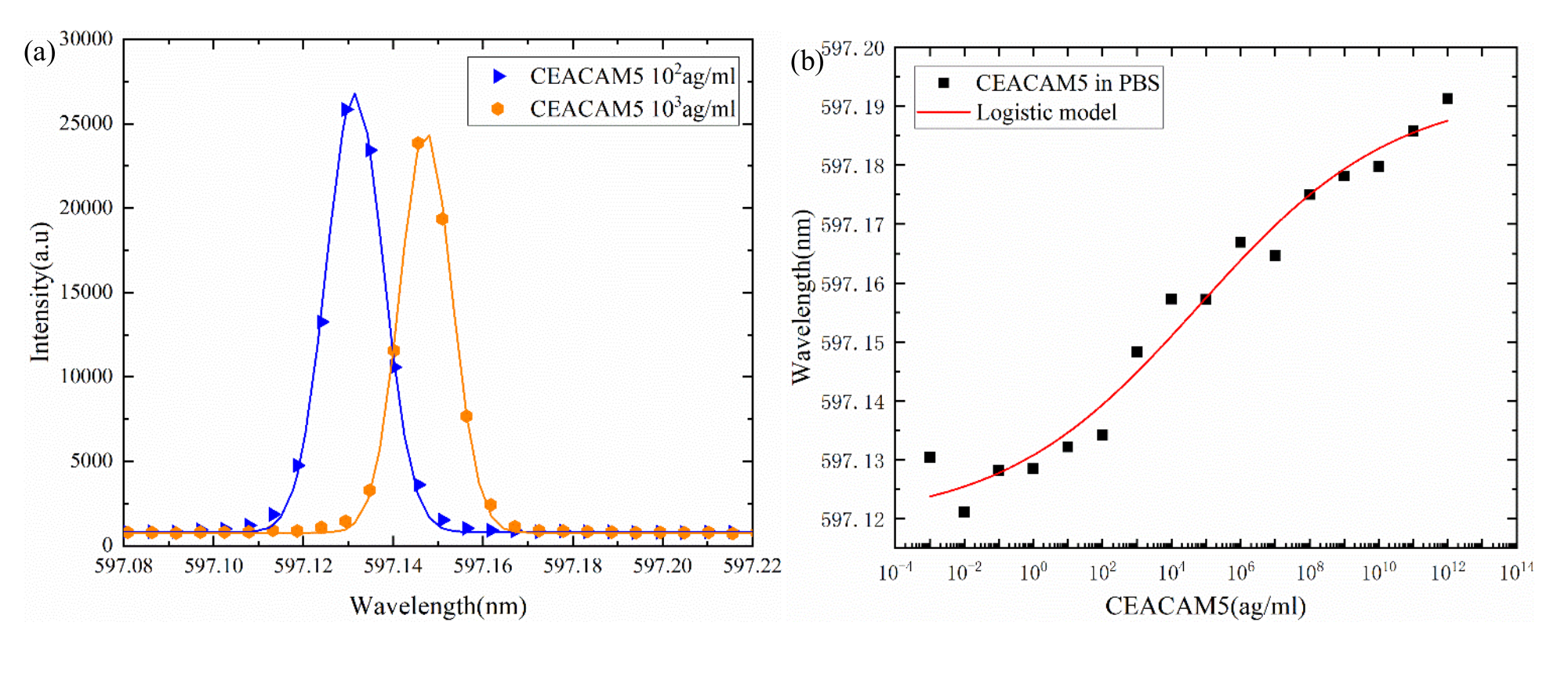}
	\caption{(a) schematic diagram of the laser biosensor. (b) Schematic diagram of the capillary resonator integrated with a microfluidic channel. (c) Microscope photo of the tapered thin-walled capillary. (d) Microscope photo of the pumped section of laser.}
    \end{figure}
	\end{document}